# Raman spectroscopy of GaSe and InSe post-transition metal chalcogenides layers

Maciej R. Molas,*[a] Anastasia V. Tyurnina,[b,c,d] Viktor Zólyomi,*[b,c] Anna K. Ott,[e,f] Daniel J. Terry,[b,c] Matthew J. Hamer,[b,c] Celal Yelgel,[c] Adam Babiński,[a] Albert G. Nasibulin,[d] Andrea C. Ferrari,[e] Vladimir I. Fal'ko,[b,c,g] Roman Gorbachev*[b,c,g]

a. Institute of Experimental Physics, Faculty of Physics, University of Warsaw, Pasteura 5, 02-093 Warszawa, Poland. Email:maciej.molas@fuw.edu.pl
b. School of Physics and Astronomy, University of Manchester, Oxford Road, M13 9PL, UK. Email: viktor.zolyomi@manchester.ac.uk, roman@manchester.ac.uk
c. National Graphene Institute, University of Manchester, Booth St E, Manchester M13 9PL, UK
d. Skolkovo Institute of Science and Technology, Nobel St. 3, 143026 Moscow, Russia
e. Cambridge Graphene Centre, University of Cambridge, 9 JJ Thomson Avenue, Cambridge CB3 0FA, UK
f. College of Engineering, Mathematics and Physical Sciences, University of Exeter, Exeter EX4 4QF, UK
g. Henry Royce Institute, University of Manchester, Oxford Road, Manchester, M13 9PL, UK

## Abstract

III-VI post-transition metal chalcogenides (InSe and GaSe) are a new class of layered semiconductors, which feature a strong variation of size and type of their band gaps as a function of number of layers (N). Here, we investigate exfoliated layers of InSe and GaSe ranging from bulk crystals down to monolayer, encapsulated in hexagonal boron nitride, using Raman spectroscopy. We present the N-dependence of both intralayer vibrations within each atomic layer, as well as of the interlayer shear and layer breathing modes. A linear chain model can be used to describe the evolution of the peak positions as a function of N, consistent with first principles calculations.

## Introduction

Layered materials (LMs) are at the center of an ever-increasing research effort spanning across a multitude of scientific disciplines,[1]. In addition to widely popular semiconducting transition metal dichalcogenides (TMDs), such as $MoS_2$, $MoSe_2$, $MoTe_2$, $WS_2$ and $WSe_2$, a new class of LMs, semiconducting post-transition metal chalcogenides (PTMCs) are increasingly studied.[2-6] These materials feature a large variation of the optical bandgap with number of layers, N: from 1.25 eV in bulk to 2.8 eV in monolayer (1L) InSe[3,4,6] and from 2.0 eV (bulk) to 2.4 eV (2L) in GaSe.[5] Furthermore, the band-gap evolves from being quasi-direct for 1L to direct in bulk. These materials also feature outstanding electronic transport properties, e.g. in InSe mobilities correspondingly up to $10^3$ $cm^2/(Vs)$ and $10^4$ $cm^2/(Vs)$ at room and liquid-helium temperatures,[4] as well as one-dimensional quantization of electrons by electrostatic gating,[7] and unusual photoluminescence, polarized primarily out of the basal plane.[6] These materials can be combined into PTMC/PTMC or PTMC/TMD layered materials heterostructures (LMHs), with type-II band alignment and allowing direct optical transitions in reciprocal space,[8] and offering an even larger selection of emission energies,[5] with potential for novel optoelectronic applications in a broad spectral range from far infra-red to violet.

GaSe and InSe crystals are anisotropic LMs comprising covalently bonded layers stacked together by van der Waals forces. Each layer consists of four atomic planes (Se-Ga-Ga-Se or Se-In-In-Se) arranged in a hexagonal atomic lattice, Figs 1a,b. In bulk, these layers can be stacked in a different orders: a hexagonal β-structure belonging to the $D_{6h}^4$ space group, a hexagonal ε-structure belonging to the $D_{3h}^1$ space group or a rhombohedral γ-structure belonging to the $D_{3v}^5$ space group.[9] However the most commonly found polytypes are ε-GaSe with a unit cell containing 8 atoms and two layers thick,[5] and γ-InSe, with a unit cell extending over 3 layers, containing 12 atoms.[6]

Raman spectroscopy is the prime non-destructive characterization tool for graphene and LMs.[10-13] In LMs there are 2 fundamentally different sets of Raman modes. Those coming from the relative motion of the atoms within each layer, usually found at high frequencies,[10,11] and those due to relative motions of the atomic planes themselves, either parallel to each other (shear or C modes)[12] or perpendicular (layer breathing modes, LBMs).[13] For the latter the experimental results are usually interpreted using with a simple linear chain model (LCM), whereby each plane is considered linked to the next one by a spring corresponding to an interlayer force constant per unit area (parallel or perpendicular to the planes for the two types of motion).[12,13]

So far, Raman characterization of PTMCs has been reported on few-layers thick crystals.[14-18] Due to low signal-to-noise ratio and strong degradation occurring in ambient conditions,[19] the thinnest PTMCs have not been investigated properly: background noise or incomplete spectra were reported, containing only one prominent Raman feature for 1L and 2L samples.[14-18]

Here, we present a Raman study of high crystalline quality InSe and GaSe flakes prepared by mechanical exfoliation, with thicknesses varying from bulk to 1L, encapsulated in hexagonal boron nitride (hBN) in an inert argon environment. We observe the thickness-dependent evolution of both intralayer vibrations as well as C and LB modes, and compare the experimental results with first principles calculations. We also demonstrate that Raman spectroscopy can be used to estimate N of PTMCs in agreement with the linear chain model.

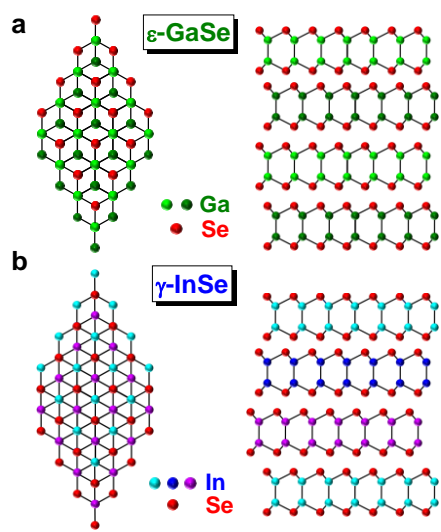

**Figure 1**: Atomic structure and stacking order of **a** ε-GaSe and **b** γ-InSe.

## Materials and methods

**Samples preparation**

GaSe and InSe samples are prepared using a specially designed micromanipulation setup placed inside an Ar filled glovebox to prevent the air-induced degradation. First, bulk PTMC crystals are mechanically exfoliated onto a 200 nm layer of poly-(propylene carbonate) (PPC) spin coated onto a bare silicon wafer and identified using optical microscopy.[20] Selected PTMC flakes are then picked up with thin hBN layers residing on a polymer membrane using a dry peel transfer technique.[19] The resulting PTMC/hBN stack is then transferred onto another hBN crystal exfoliated onto a Si/SiO$_2$ substrate to achieve full encapsulation. The verification of the PTMC thicknesses is performed using atomic force microscopy (AFM).

**Experimental setups**

Raman measurements are carried out in backscattered geometry at room temperature using Renishaw, Horiba LabRam HR Evolution, Horiba XploRa and Horiba FHR1000 systems equipped with interference or Bragg grating filters with a cut-off at 5 cm$^{-1}$. The Raman spectra are taken 633 nm (1.96 eV), 532 nm (2.33 eV), 515 nm (2.41 eV), 488 nm (2.54 eV), 457 nm (2.71 eV). The laser is focused by a 100x objective. The laser power on the sample is kept at~150 μW during all measurements, to avoid laser induced damage and local heating. The collected signal is dispersed with 1800 and 2400 grooves mm$^{-1}$ gratings and detected with a CCD camera. For each measurements, the spectrum from an area near each studied layer is subtracted, to remove the background signal from the Si substrate. The background spectra are also used for accurate calibration of the data from different setups using the Si Raman mode at 520.7 cm$^{-1}$.[21]

## Experimental results and discussion

Fig.2 plots the polarization-resolved Raman spectra of 3L GaSe and InSe, accompanied by the corresponding atomic displacement of the phonon modes. 3L GaSe has two out-of-plane phonon modes, $A'_1(1)$ at ~135 cm$^{-1}$ and $A'_1(2)$ at ~308 cm$^{-1}$, and one in-plane $E''(2)$ mode at ~214 cm$^{-1}$. 3L InSe has 3 out-of-plane modes: $A'_1(1)$ at ~115 cm$^{-1}$, $A''_2(1)$ at ~200 cm$^{-1}$ and $A'_1(2)$ at ~228 cm$^{-1}$, and one in-plane $E''(2)$ mode at ~178 cm$^{-1}$. The peaks are classified according to their irreducible representation in the symmetry group in the 1L phase D$_{3h}$ and additionally numbered due to their increased Raman shift. This notation is consistent with earlier works on bulk ε-GaSe, γ-InSe.[22-25]

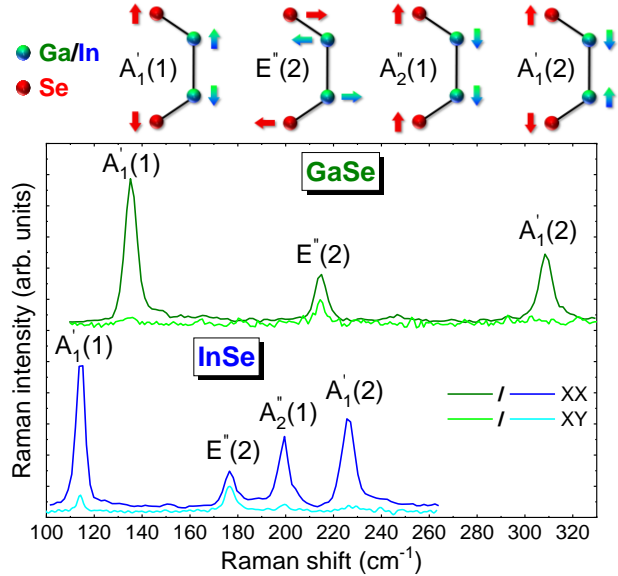

**Figure 2**: Atomic displacements of phonon modes and polarization-resolved 514.5 nm Raman spectra of 3L GaSe (green/bright green) and InSe (blue/bright blue) in XX/XY configurations.

To confirm their symmetries, the Raman spectra are recorded at 488 nm for 2 polarizations: parallel – XX and cross-polarized – XY. For both materials, all peaks are observed in XX, while only $E''(2)$ modes are seen in XY, as expected due to the rotational symmetry in these crystals, which implies that A-type modes are only active in parallel polarization, while E-type modes are observable in both (see Supplementary Information for details).

In order to reveal evolution of Raman fingerprint with the layer thickness, crystals with multiple terraces with lateral sized >1 μm have been studied. The unpolarized Raman spectra, presented in Figs. 3a,b, show the optical phonon modes down to 1L.

For GaSe, $A'_1(1)$ and $A'_1(2)$ demonstrate red- and blue-shift, respectively, with the decreasing N, while $E''(2)$ is barely affected by N. For bulk, a double peak is observed, comprising the $E''(2)$ at ~211 cm$^{-1}$ and $E'(2)$ at ~216 cm$^{-1}$, consistent with first principles calculations (see Supplementary Information), and previous Raman studies of bulk crystals.[22,23] The $E'(2)$ mode corresponds to an in-plane phonon mode, where Ga or Se vibrate in-phase, while an out-of-phase vibration occurs between Ga and Se pairs.

For InSe, Fig. 3b, we also observe a similar red- and blue-shift of $A'_1(1)$ and $A'_1(2)$ with decreasing N, and no significant shift of $E''(2)$, while $A''_2(1)$ blueshifts like $A'_1(2)$. For 16L and bulk, an additional peak is seen ~210 cm$^{-1}$, visible under XY polarization, suggesting to be an E-type vibration. As for Refs. 26-28, we assign it to an $E'(2)$ mode, similar to GaSe bulk. The observation of $A''_2(1)$ in the Raman spectrum of InSe bulk was previously ascribed to resonance with the B excitonic transition.[25] Ref. 3 showed that the B energy is substantially affected by N, changing from ~2.4 eV in bulk to ~2.9 eV for 1L. For a constant excitation of 2.54 eV as used in Fig.3, the resonance is strongest for intermediate thicknesses, with the largest $A''_2(1)$ intensity for 5L, Fig. 3b.

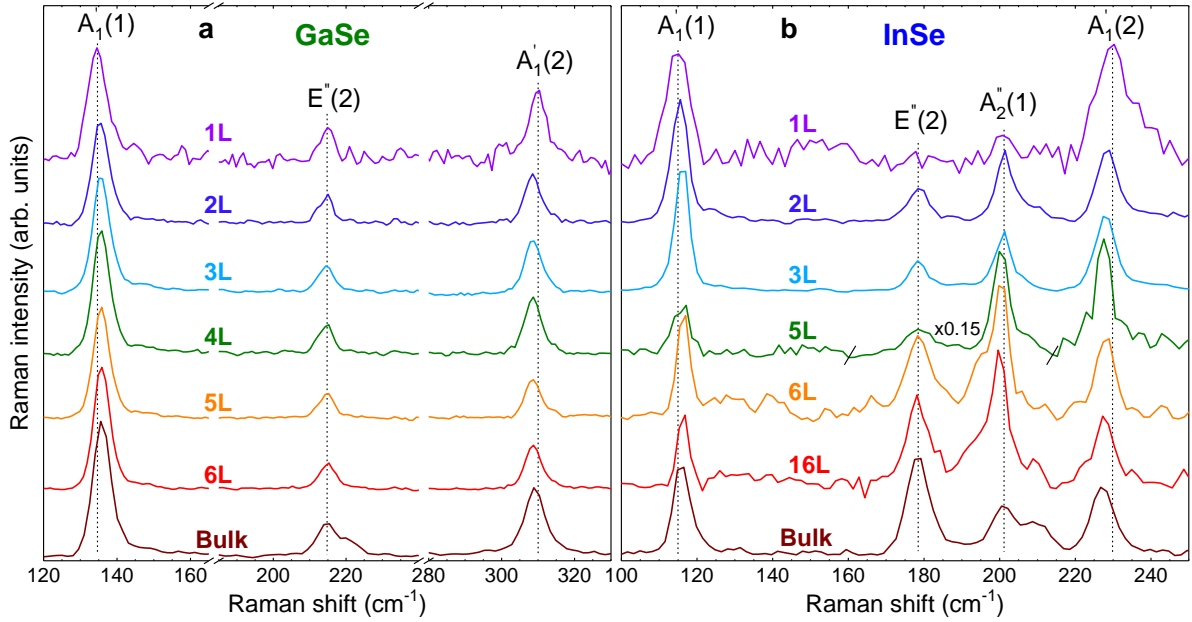

**Figure 3:** 488nm unpolarized Raman as a function of N for **a** GaSe and **b** InSe. Dashed lines show peak positions for 1L.

The summary of the peak position evolution with N, extracted using Lorentzian fitting of the Raman spectra for both materials is shown in Figs.4a,b. Previous publications have reported only one phonon mode in thin crystals down to 1L, i.e. $A'_1(2)$ for GaSe[15] and $A'_1(1)$ for InSe,[16] likely due to the lack of encapsulation, leading to significant crystal degradation.[29] The reported N evolution of $A'_1(2)$ in GaSe[15] is opposite (a redshift with decreasing N) to our observations, which may be due to an overlap of the $A'_1(2)$ mode with Si modes ~300 cm$^{-1}$ in Ref. 20. In InSe, a similar redshift of $A'_1(1)$ was previously reported down to 2L.[16] However, the Raman signal of 1L was not detected and only $A'_1(1)$ was visible in 2L.

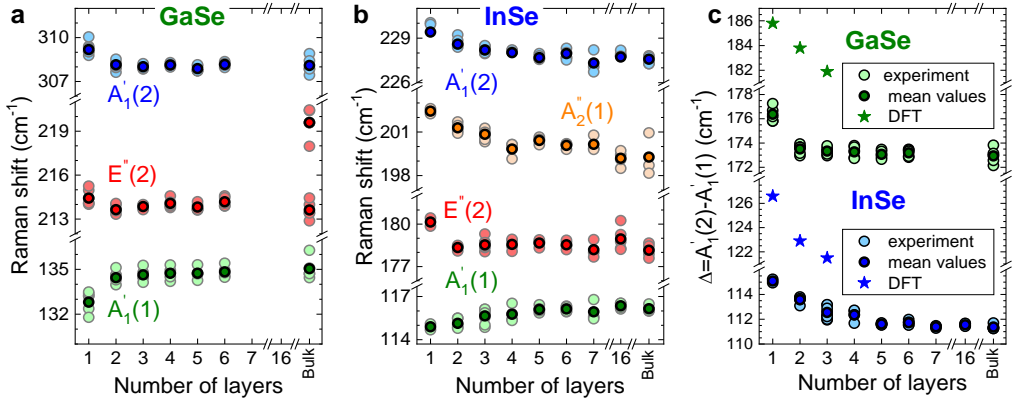

**Figure 4:** N-dependent Raman shifts for **a** GaSe and **b** InSe. **c** N dependence of $\Delta = A'_1(2) - A'_1(1)$. DFT calculations are indicated by star symbols. The dark color symbols correspond to mean values for the dataset shown in light colors measured at 633, 533, 514.5, 488 and 457 nm.

We further investigate the N dependence of the difference in peak position between $A'_1(2)$ and $A'_1(1)$, i.e. $\Delta = A'_1(2) - A'_1(1)$. According to our DFT calculations presented in Supplementary Information, a gradual rise in $\Delta$ with decreasing N is predicted for both materials, Fig.4c. This matches our experimental results for InSe, where $\Delta$ is gradually increasing from 5L to 1L, with a total blueshift ~3.5 cm$^{-1}$. A more abrupt change is observed in GaSe, where $\Delta$ only changes noticeably in 1L by ~3 cm$^{-1}$. $\Delta$ in both materials is substantial enough to be used for assessing N in FL-InSe and 1L-GaSe, similar to what done for TMDs.[29,30]

Fig. 5 plots the low frequency Raman spectra of InSe and GaSe flakes. Figs. 5a,b show 3 types of Raman peaks for both materials. Two of them are due to the vibrations in the GaSe and InSe layers, while the peak ~52 cm$^{-1}$, indicated by * is the C peak of the encapsulating hBN.

The evolution of C and LB modes can be described using a linear chain model (LCM), where the layers are represented as one mass connected to the next layer by a spring with a coupling constant. The $Pos(C)$ and $Pos(LBM)$ can then be written as[11-13]:

$$Pos\left(C_{N,N-i}\right) = \frac{1}{\sqrt{2}\pi c}\sqrt{\frac{\alpha_{\parallel}}{\mu}}\sin\left[\frac{i\pi}{2N}\right] \quad (1)$$

$$Pos\left(LBM_{N,N-i}\right) = \frac{1}{\sqrt{2}\pi c}\sqrt{\frac{\alpha_{\perp}}{\mu}}\sin\left[\frac{i\pi}{2N}\right] \quad (2)$$

where $\alpha_{\parallel}$ and $\alpha_{\perp}$ are force constants per unit area for the shear motion, denoted ∥ and the layer breathing motion, denoted ⊥, between two layers. $\mu$ is the 1L mass per unit area. *i=1,2,…N-1*. Multiplying $\alpha_{\parallel}$ and $\alpha_{\perp}$ by the unit cell area gives the interlayer force constants, $K$.[11-13]

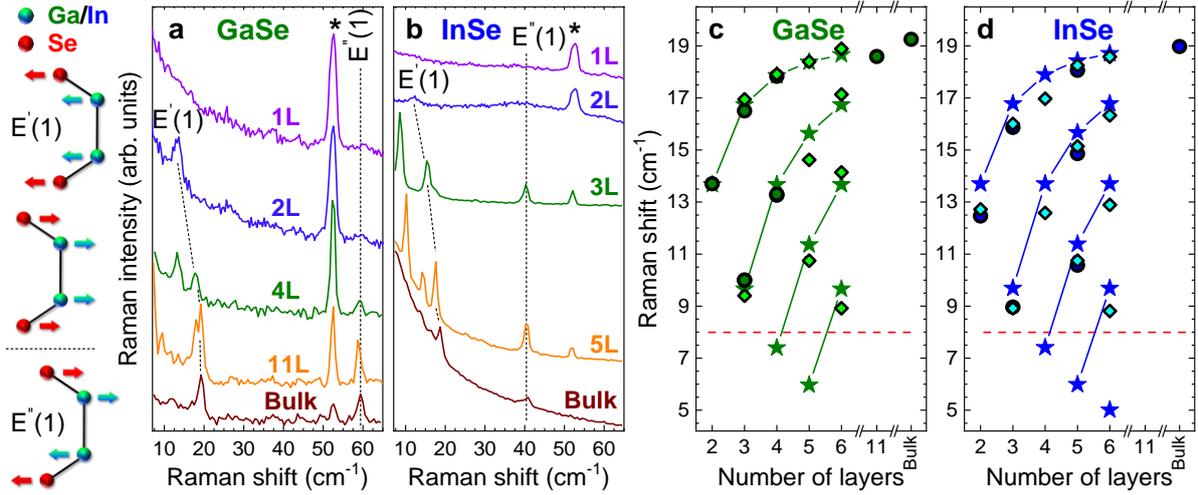

**Figure 5**: Atomic displacements of phonon modes and unpolarized 514.5 nm low-frequency Raman spectra as a function of N for **a** GaSe and **b** InSe. Dashed lines are guides to the eye. (c,d) N dependence of C modes of **c** GaSe and **d** InSe. Circles and diamonds denote experimental results (for diamonds data is not shown). Stars are DFT calculations. Dashed red horizontal lines represent the minimal Raman shift ~8 cm$^{-1}$ possible to detect in experiments

We also use DFT within the local density approximation (LDA) to calculate the vibrational frequencies, as detailed in the SI. To compute the C and LBM frequencies, we rely on the frozen phonon approximation to calculate the corresponding force constant between neighboring layers. The interlayer force constant remains the same within 1% when increasing N from 2 to 3. Therefore, the LCM approach can be applied to C modes under the assumption that independent force constants describe the coupling between neighboring layers. Such an approximation was previously applied to ε-GaSe.[32] We obtain the interlayer force constants: for C modes, $K^{\text{GaSe}} = 102$ meV/Å$^2$ and $K^{\text{InSe}} = 134$ meV/Å$^2$; for LBMs, $K^{\text{GaSe}} = 323$ meV/Å$^2$ and $K^{\text{InSe}} = 401$ meV/Å$^2$.

In experiment, we only observe C modes. Fig. 5c,d plot experimental and calculated Pos(C). The results are in good agreement, both concerning frequencies and number of branches for a given N. By fitting the measured Pos(C) as function of N with Eq. 1, we get $\alpha_{\parallel}$ ~7.32·10$^{18}$ N/m for InSe and ~6.72·10$^{18}$ N/m for GaSe. We use the experimental values for the lattice constants $a_{InSe}^{exp} = 4.002$ Å[33] and $a_{InSe}^{exp} = 3.755$ Å[34] to extract the unit area. Using the LDA lattice parameters $a_{GaSe}$=3.6678 Å and $a_{InSe}$=3.9103 Å and computing the unit cell areas, we can extract $\alpha_{\parallel}$ from the theoretical interlayer force constants $K$ for InSe and GaSe by dividing $K$ by the unit cell area. We get 1.01·10$^{19}$N/m for InSe and 8.76·10$^{18}$N/m for GaSe, in good agreement with the experimentally determined values.

## Conclusions

We have performed a Raman spectroscopic investigation of InSe and GaSe flakes of varying number of layers from the bulk limit down to a monolayer, for samples encapsulated in hBN. We determined the thickness-dependent evolution of both intralayer and interlayer modes, and compared the results with first principles calculations. We confirmed that Raman spectroscopy can be used to estimate the number of layers of PTMCs.

## Conflicts of interest

There are no conflicts to declare.

## Acknowledgements

We acknowledge support from the National Science Centre, Poland (grants no. 2017/24/C/ST3/00119, 2017/27/B/ST3/00205), EPSRC ARCHER RAP grant (e547), Graphene Flagship, Quantum Technology Flagship (820378), The Scientific and Technological Research Council of Turkey (TUBITAK) through BIDEB-2219 programme (2018-1), EPSRC CDT Graphene-NOWNANO and Graphene Technology, EPSRC Doctoral Prize Fellowship, and a Royal Society Research Fellowship Scheme, Samsung Advanced Institute of Technology (SAIT), ERC Grant Hetero2D, EPSRC Grants EP/K01711X/1, EP/K017144/1, EP/N010345/1and EP/L016057/1.